\documentclass[twocolumn,prl,floatfix,superscriptaddress,nofootinbib]{revtex4-1}

\usepackage[pdftex]{graphicx}
\usepackage{amsmath,amssymb,bm}

\graphicspath{{./Figures/}}

\begin{document}
	
\title{Search for axion-like dark matter with ferromagnets}

\author{Alexander~V.~Gramolin}
\affiliation{Department of Physics, Boston University, Boston, MA 02215, USA}
\author{Deniz~Aybas}
\affiliation{Department of Electrical and Computer Engineering, Boston University,Boston, MA 02215, USA}
\author{Dorian~Johnson}
\affiliation{Department of Physics, Boston University, Boston, MA 02215, USA}
\author{Janos~Adam}
\affiliation{Department of Physics, Boston University, Boston, MA 02215, USA}
\author{Alexander~O.~Sushkov}
\email{asu@bu.edu}
\affiliation{Department of Physics, Boston University, Boston, MA 02215, USA}
\affiliation{Department of Electrical and Computer Engineering, Boston University,Boston, MA 02215, USA}
\affiliation{Photonics Center, Boston University, Boston, MA 02215, USA}

\begin{abstract}
	Existence of dark matter indicates the presence of unknown fundamental laws of nature. Ultralight axion-like particles are well-motivated dark matter candidates, emerging naturally from theories of physics at ultrahigh energies. We report the results of a direct search for the electromagnetic interaction of axion-like dark matter in the mass range that spans three decades from 12~peV to 12~neV. The detection scheme is based on a modification of Maxwell's equations in the presence of axion-like dark matter, which mixes with a static magnetic field to produce an oscillating magnetic field. The experiment makes use of toroidal magnets with iron-nickel alloy ferromagnetic powder cores, which enhance the static magnetic field by a factor of~24. Using SQUIDs, we achieve a magnetic sensitivity of 150~$\text{aT}/\sqrt{\text{Hz}}$, at the level of the most sensitive magnetic field measurements demonstrated with any broadband sensor. We recorded 41 hours of data and improved the best limits on the magnitude of the axion-like dark matter electromagnetic coupling constant over part of our mass range, at 20~peV reaching $4.0 \times 10^{-11}~\text{GeV}^{-1}$ (95\% confidence level). Our measurements are starting to explore the coupling strengths and masses of axion-like particles where mixing with photons could explain the anomalous transparency of the universe to TeV gamma-rays.
\end{abstract}

\maketitle

Astronomical evidence, built up over eight decades, indicates that only one-sixth of matter in the universe is made up of fundamental particles whose properties we understand~\cite{Spergel2015, PDG2019}. The existence of dark matter constitutes compelling evidence for new fundamental particles and interactions beyond the Standard Model. Since it has only been observed through its gravitational effects, there is a broad range of dark matter candidates. The weakly-interacting massive particle (WIMP) is a candidate that has inspired a large number of ultra-sensitive experiments of increasing complexity and scale, but so far there has been no unambiguous detection, and fundamental backgrounds (neutrino floor) will soon start to limit the sensitivity of direct WIMP searches~\cite{Liu2017, PDG2019, Rajendran2017}. Another well-motivated dark matter candidate is the axion, which also offers a compelling solution to the strong CP problem of quantum chromodynamics~\cite{Preskill1983, Abbott1983, Dine1983, DeMille2017, Irastorza2018a}. The axion and other light pseudoscalar bosons (axion-like particles, ALPs) emerge naturally from theoretical models of physics at high energies, including string theory, grand unified theories, and models with extra dimensions~\cite{Svrcek2006, Irastorza2018a}. Astrophysical observations have produced a number of stringent limits on ALPs, but also some tantalizing hints for their possible existence~\cite{PDG2019}. One such hint is the observation that the universe is too transparent to TeV $\gamma$-rays. These are emitted from distant active galactic nuclei and should be significantly attenuated, through pair production, by infrared background radiation during their travel to our galaxy~\cite{Matsuura2017}. The tension with observed TeV gamma-ray source energy spectra can be explained by photon-ALP mixing inducing gamma-ray conversion into axions, which travel unaffected through interstellar space and convert back into gamma-rays in the Milky Way~\cite{Kohri2017}.

Experimental searches for the axion and ALPs rely on one of their interactions with Standard Model particles~\cite{Graham2013, Budker2014, Arvanitaki2014, Irastorza2018a}. Most experiments to date have focused on the axion-photon interaction, which can mix photons with axions and ALPs in the presence of a strong magnetic field. The conversion rate scales with magnetic field and volume, so experiments employ large magnetic fields and volumes for better sensitivity~\cite{Battesti2018}. This effect has been used to place stringent limits on the coupling of ALPs produced in the laboratory~\cite{Ehret2010}, or in the Sun~\cite{Anastassopoulos2017}. Several searches for axion dark matter in the $\mu$eV mass range have used the technique of resonant conversion into monochromatic microwave photons inside a high-quality-factor cavity permeated by a strong magnetic field~\cite{Graham2015a}. The ADMX experiment has achieved a level of sensitivity sufficient to search for dark matter axions with masses between 2.66 and $2.81~\mu\text{eV}$ and excluded the QCD axion-photon couplings predicted by plausible models for this mass range~\cite{Du2018}. A number of cavity-based axion dark matter searches are in development or already exploring ALP masses up to $\approx 30~\mu\text{eV}$~\cite{Graham2015a, Brubaker2017, Choi2017b}. In order to search for lower-mass axions and ALPs coupled to photons, it is possible to use lumped-element circuits~\cite{Sikivie2014b, Chaudhuri2015, Kahn2016, Chaudhuri2018, Ouellet2019}. This concept is based on a modification of Maxwell's equations: in the presence of a large static magnetic field $B$, axion-like dark matter acts as a source of an oscillating magnetic field whose amplitude is proportional to~$B$~\cite{Sikivie1983, Wilczek1987}.

Our approach is to use a toroidally-shaped permeable material to enhance the magnitude of the static magnetic field~$B$ and thus improve sensitivity to axion-like dark matter. In the presence of a magnetizable medium, the modified inhomogeneous magnetic Maxwell's equation takes the form 
\begin{equation}
\vec{\nabla}\times\vec{H} = \vec{J}_f + \frac{g_{a\gamma}}{\mu_0c}\frac{\partial a}{\partial t}\vec{B},
\label{eq:1}
\end{equation}

\begin{figure*}[t]
	\centering
	\includegraphics[width=0.8\textwidth]{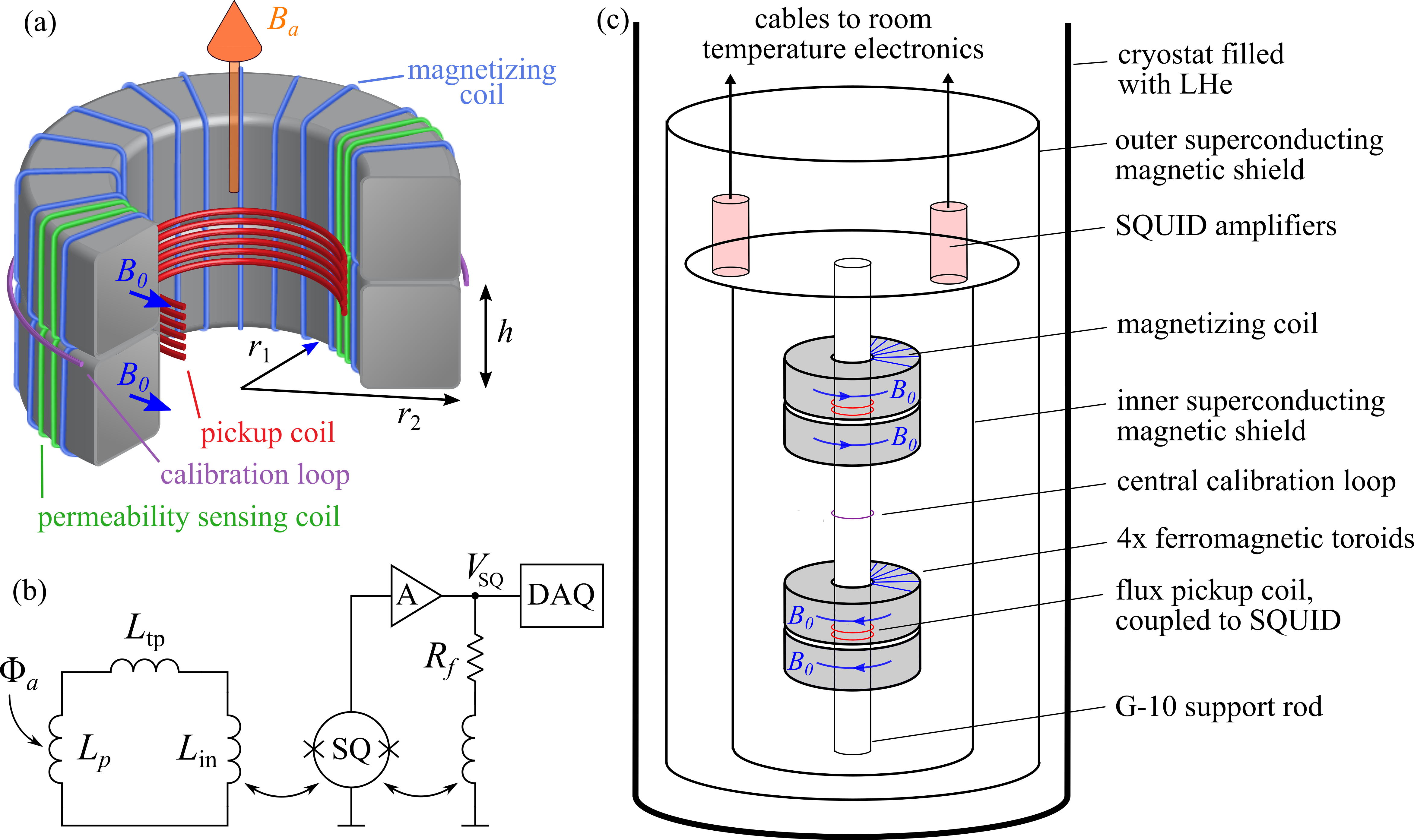}
	\caption{
		(a) Schematic of a single detection channel. Two permeable toroids were independently magnetized by injecting current into a magnetizing coil wrapped around each toroid. The electromagnetic coupling~$g_{a\gamma}$ of axion-like dark matter induces an oscillating magnetic field $B_a$, which produces a magnetic flux $\Phi_a$ through the pickup coil mounted at the inner circumference of the toroids. Toroid magnetization was monitored by measuring the inductance of the permeability sensing coils. Dimensions: $r_1 = 24.4$~mm, $r_2 = 39.1$~mm, $h = 16.2$~mm.
		(b) The circuit model that shows the axion-induced flux~$\Phi_a$ coupling into the pickup coil (inductance $L_p$). The pickup coil was coupled to the SQUID magnetic flux sensor (SQ) via twisted pair leads (inductance $L_{\rm tp}$) and input coil (inductance $L_{\rm in}$). The SQUID was operated in the flux-locked-loop mode, with feedback resistance $R_f$. The feedback voltage $V_{\rm SQ}$ was digitized and recorded by the data acquisition system (DAQ)~\cite{som}.
		(c) The SHAFT experimental schematic. The apparatus contained four permeable toroids, shown magnetized in the ($+ +$, $- -$) configuration, which was used to collect data sensitive to axion-like dark matter. The experiment operated at 4.2~K in a liquid helium bath cryostat.
	}
	\label{fig:1}
\end{figure*}

\noindent
where $\vec{H} = \vec{B}/\mu_0 - \vec{M}$ is the auxiliary field, $\vec{M}$ is the magnetization, $\vec{J}_f$ is the macroscopic free electric current density, $\mu_0$ is the permeability of free space, $c$ is the speed of light in vacuum, and $g_{a\gamma}$ is the strength of ALP electromagnetic coupling~\cite{som}. The axion-like dark matter field $a = a_0 \sin{(\omega_a t)}$ oscillates at its Compton frequency $\nu_a=\omega_a/(2\pi)=m_ac^2/h$, where $m_a$ is the ALP mass, and $h$ is the Planck constant. The coherence time of this oscillating field is $\approx 10^6$ periods, set by the kinetic energy of the virialized axionic dark matter~\cite{Sikivie1983, Graham2013}. We use SI units for electromagnetic fields and natural units for the ALP field so that $g_{a\gamma} a_0$ is unitless, and the ALP field amplitude~$a_0$ is given by the dark matter energy density: $m_a^2 a_0^2 / 2 = \rho_{\text{DM}} = 3.6 \times 10^{-42}~\text{GeV}^4$~\cite{PDG2019, Graham2013}.

Our experiment, the Search for Halo Axions with Ferromagnetic Toroids (SHAFT), contained two independent detection channels. Each channel consisted of two stacked toroids, each of which could be independently magnetized by injecting a current into a superconducting magnetizing coil wound around the toroid, fig.~\ref{fig:1}(a). With azimuthal field $B_0$ inside the magnetized toroid, the second term in the right-hand side of eq.~(\ref{eq:1}) represents an effective current density, $J_{\rm eff} = \omega_a g_{a\gamma} a_0 \cos{(\omega_a t)} B_0 / (\mu_0 c)$, sourced by the axion-like dark matter field. This loop of current creates an axial magnetic field $B_a$. The pickup coil, wound around the inner circumference of the toroids, was coupled to a Superconducting Quantum Interference Device (SQUID) magnetometer, as shown in fig.~\ref{fig:1}(b), measuring the magnetic flux~$\Phi_a$ due to~$B_a$. The axion-like dark matter detection signature would be an oscillating SQUID output signal, whose amplitude was correlated with toroid magnetization.

\begin{figure*}[t]
	\centering
	\includegraphics[width=0.8\textwidth]{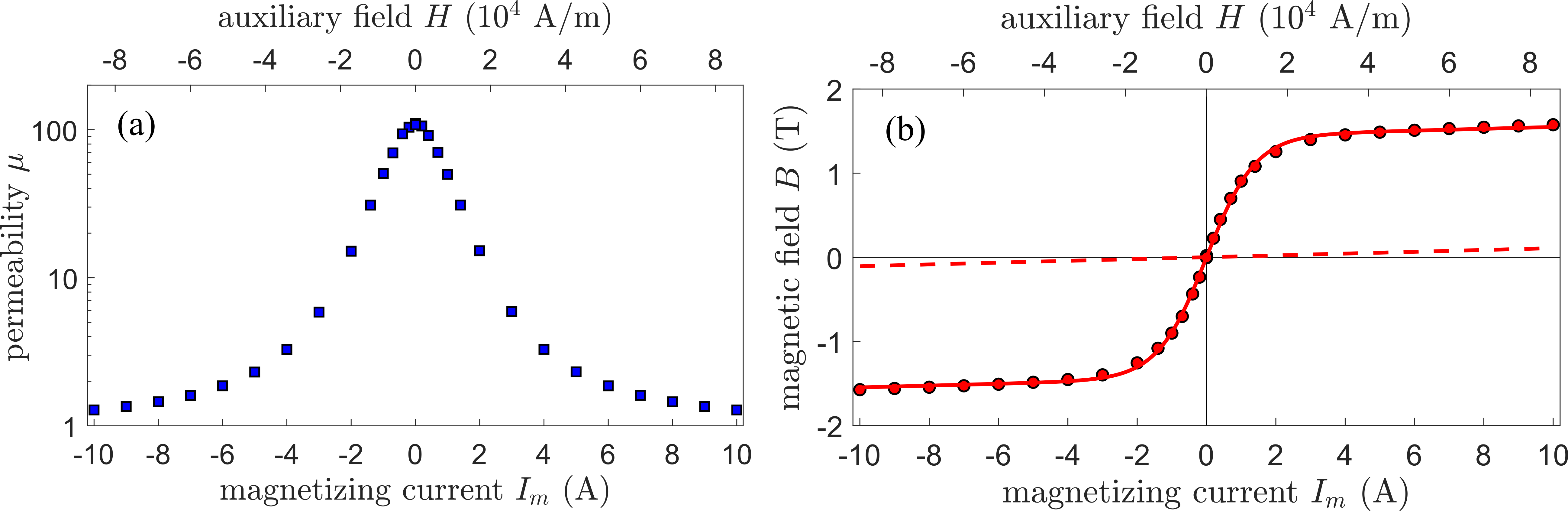}
	\caption{
		(a) Measurements of the toroid permeability at temperature 4.2~K, for currents up to 10~A injected into the magnetizing coil. Permeability decreases with increasing current, as the magnetic material saturates.
		(b) The azimuthal magnetic field $B_0$ inside the toroid, calculated by integrating the permeability data. Solid line shows the saturation model: $B = \mu_0 \left[H + M_0 \tanh(H/H_0)\right]$, with best-fit saturation magnetization $M_0=(1.15\pm 0.03)\times 10^6$~A/m and saturation field $H_0=(1.20\pm 0.05)\times 10^4$~A/m~\cite{som}. Dashed line shows the magnetic field that would be created inside an air-core toroid.
	}
	\label{fig:2}
\end{figure*}

The two-channel, four-toroid design enabled a systematic rejection scheme that allowed for discrimination between an axion-like dark mater signal and electromagnetic interference, fig.~\ref{fig:1}(c). For collecting data sensitive to axion-like dark matter, we magnetized the top two toroids counter-clockwise (channel A: $+ +$), and the bottom two toroids clockwise (channel B: $- -$), viewed from the top. An axion-like dark matter signal would then appear in the two detector channels with opposite sign 
($180^{\circ}$ out of phase). Electromagnetic interference, uncorrelated with toroid magnetization, would not have such a phase relationship and could be distinguished from an axion-like dark matter signal by forming symmetric and antisymmetric combinations of the two detector channels. The toroids and pickup coils were inside an enclosure formed by two nested coaxial cylinders immersed in liquid helium. Lead foil, affixed to the inner surfaces of the cylinders and their caps, formed a double-layer superconducting magnetic shield, suppressing electromagnetic interference and ambient magnetic field noise.

The sensitivity enhancement in our approach arises from the fact that the magnetic field inside the permeable toroids includes material magnetization in addition to the field created by the free current in the toroidal magnetizing coil: $\vec{B}_0 = \mu_0 (\vec{H}_0 + \vec{M}_0)$. For a linear magnetic material with permeability $\mu$, $B_0 = \mu \mu_0 H_0$ and the magnetic field enhancement is by a factor equal to $\mu$. In practice, magnetic materials are non-linear: the permeability drops with applied field $H_0$ and the magnetization saturates. For our experiment, we chose powdered iron-nickel ``High Flux $125\mu$'' alloy for its high permeability, high saturation flux density, and electrical insulating properties at liquid helium temperature~\cite{som}. We determined the permeability of the toroid material by injecting currents up to 10~A into the magnetizing coil and measuring its inductance. The initial permeability was $\approx 110$, dropping with increasing coil current as the material saturated, fig.~\ref{fig:2}(a). The material displayed no magnetic hysteresis. We used the permeability measurements to calculate magnetic field $B_0$ inside the toroid at each value of the magnetizing coil current using $B_0(H_0) = \mu_0 \int_0^{H_0} \mu(H) \, dH$, fig.~\ref{fig:2}(b). Before recording data sensitive to ALP dark matter, we set the magnetizing coil current to 6~A, at which point $B_0 = 1.51~\text{T}$ and the material is close to saturation. At this current, the magnetic field inside an air-core toroid would be 0.063~T (dashed red line in fig.~\ref{fig:2}(b)). Therefore, we can quantify the enhancement of $B_0$ due to the magnetic material to be a factor of 24. During ALP-sensitive data collection, we switched each of the magnetizing coils into persistent mode, but still monitored each toroid magnetization by making inductance measurements of separate permeability sensing coils, which had been calibrated in advance~\cite{som}.

\begin{figure*}[t]
	\centering
	\includegraphics[width=0.7\textwidth]{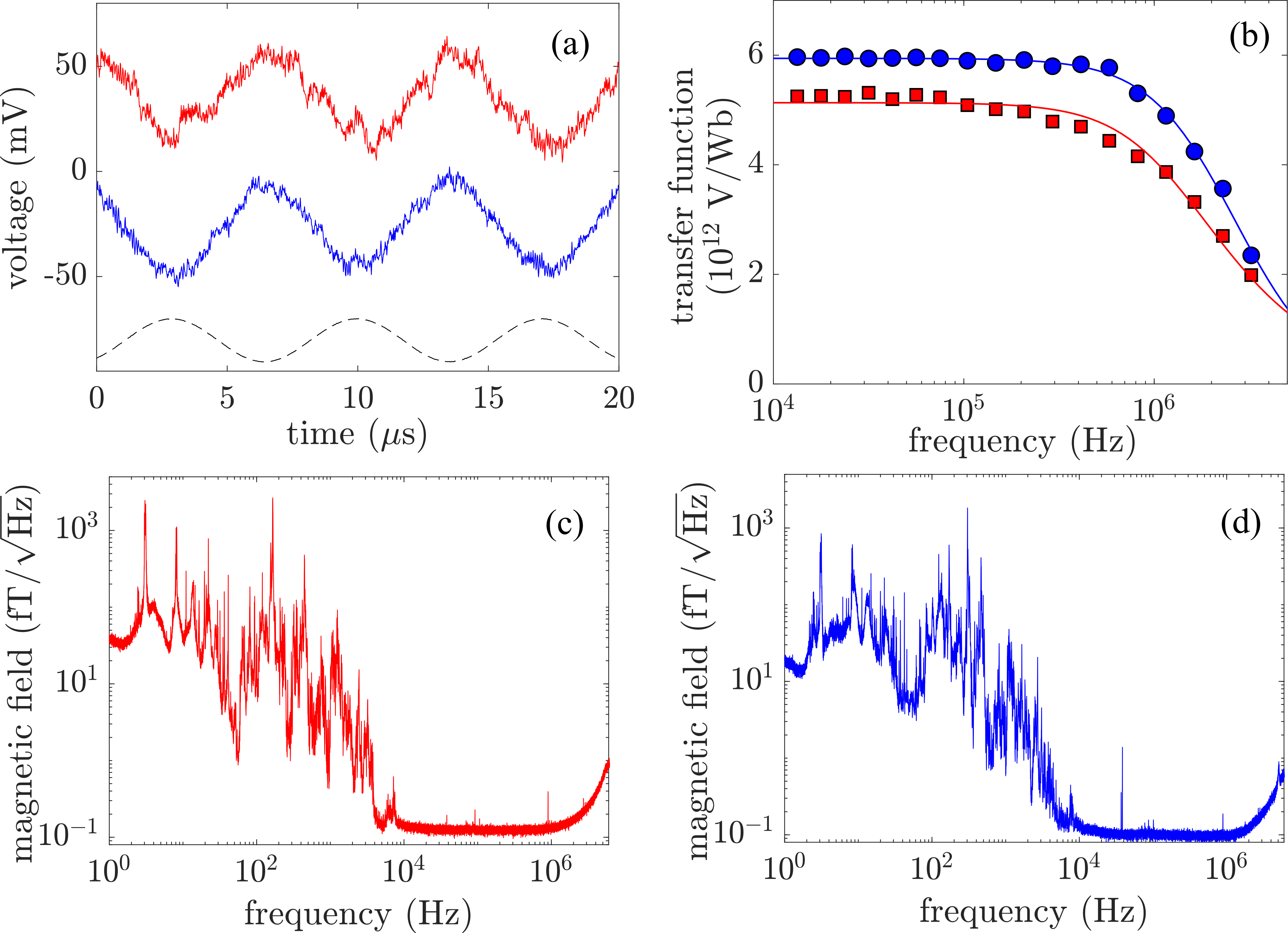}
	\caption{
		(a) Magnetic flux calibration measurements: sinusoidally-varying current was injected into the central calibration loop and the response of the two SQUID detection channels was recorded. Dashed black line shows the signal proportional to injected current at 142~kHz; red and blue lines show the time-domain response of channels A and B, respectively (offset along the y-axis for clarity). At this frequency, the channels are in phase. The full quadrature calibration was performed as a function of excitation frequency up to 4~MHz~\cite{som}.
		(b) The frequency dependence of the absolute value of SQUID flux-to-voltage transfer function for channel A (red squares) and channel B (blue circles). The low-frequency value of the transfer function was consistent with the inductive coupling model shown in fig.~\ref{fig:1}(b), and the high-frequency roll-off was due to the finite bandwidth of the SQUID flux-locked loop feedback electronics.
		(c,d) SQUID magnetic sensor noise spectra for channels A and B, respectively. Noise at frequencies below 10~kHz was correlated with cryostat vibrations, and the degradation at frequencies above 1~MHz was due to SQUID transfer function roll-off.
	}
	\label{fig:3}
\end{figure*}

Quantifying the sensitivity of our experiment to an axion-like dark matter signal required measurements of the magnetic field sensitivity of the two detection channels. The experimental design included four independent single-turn calibration loops, mounted at several positions near the toroids~\cite{som}. We performed calibration measurements by injecting sinusoidally-varying current with known frequency and amplitude into each loop and recording the response of both SQUID channels. Measurements with the central calibration loop, which was positioned midway between the upper and the lower toroid pairs, and thus had equal couplings to the two pickup coils, showed that the detector channels were in phase, fig.~\ref{fig:3}(a). We recorded the SQUID output~$V_{\rm SQ}$ for a range of calibration current frequencies and used the numerically-computed mutual inductance to calculate the magnetic flux $\Phi$ through each pickup coil. These measurements established the calibration of the flux-to-voltage transfer function $x = V_{\rm SQ}/\Phi$, whose magnitude was consistent with the inductive coupling model shown in fig.~\ref{fig:1}(b), and whose frequency dependence was consistent with a second-order low-pass filter model, fig.~\ref{fig:3}(b). The transfer function measurements for all calibration loops mounted inside the experiment were consistent within a $25\%$ range~\cite{som}.

The main factor limiting the sensitivity of our ALP dark matter search was SQUID sensor noise. To quantify magnetic sensitivity, we magnetized the toroids with 6~A injected current, recorded the SQUID output voltage, converted it to magnetic field using the transfer function for each SQUID detector channel, and calculated the Fourier spectra, fig.~\ref{fig:3}(c,d). Below 3~kHz, the spectra were dominated by vibrations of the pickup coil in the magnetic field leaking out of the magnetized toroids, and above 3~MHz, the sensitivity was degraded by the SQUID feedback electronics bandwidth roll-off. Aside from several noise peaks due to residual radiofrequency (RF) interference, between 10~kHz and 1~MHz the noise spectrum was very nearly flat at the level of $150$~aT/$\sqrt{\rm Hz}$. This is at the level of the most sensitive magnetic field measurements demonstrated with broadband SQUID systems~\cite{Storm2017} or atomic magnetometers~\cite{Dang2010}, in spite of the direct proximity of ferromagnetic material magnetized to 1.51~T. The absence of detectable magnetization noise from the ferromagnetic toroids was due to the pickup coil geometry, designed to be insensitive to toroid magnetization fluctuations~\cite{Eckel2009, Sushkov2009}. For our axion-like dark matter search, we chose the frequency range between 3~kHz and 3~MHz so that the noise was within a factor of~3 of the baseline.

\begin{figure*}[t]
	\centering
	\includegraphics[width=0.7\textwidth]{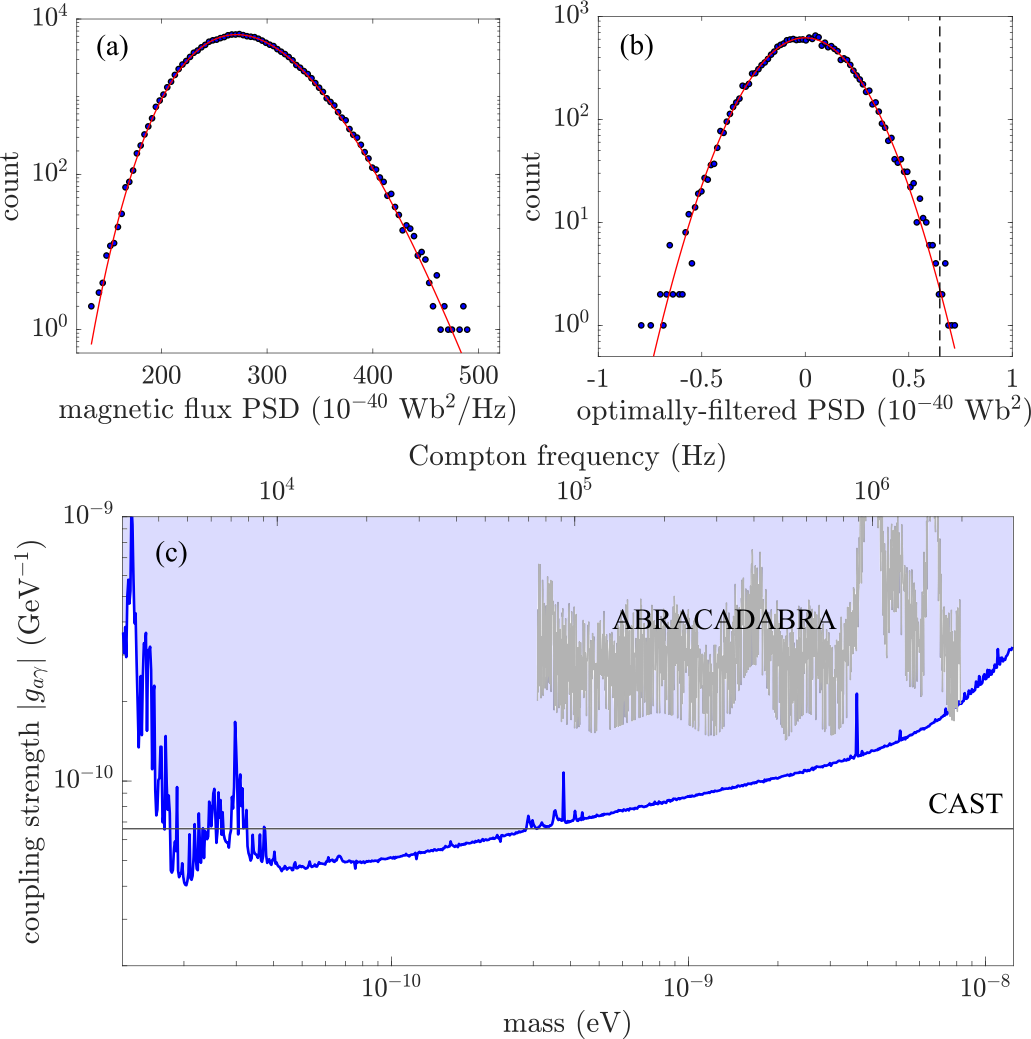}
	\newpage
	\caption{
		(a) The histogram of the magnetic flux power spectral density data within a frequency bin centered at the frequency $\nu_b = 10.04$~kHz and with the width $10^4 \beta^2 \nu_b \approx 0.08$~kHz. The histogram is consistent with a chi-square distribution with 94 degrees of freedom (red line). This distribution is expected because the PSD is formed by adding the squares of the real and imaginary parts of the Fourier transform and averaging over 47 data blocks. Therefore, it is a sum of squares of 94 independent normally-distributed random variables~\cite{Walck2007}.
		(b) The histogram of data in the $\nu_b = 10.04$~kHz bin, optimally-filtered by convolving with the standard halo model axion-like dark matter signal spectral shape. The red line shows the normal distribution fit, and the dashed vertical line shows the $3.355\sigma$ threshold for flagging candidate frequencies. 
		(c) Limits on the electromagnetic coupling strength of axion-like dark matter in the mass range from 12~peV to 12~neV. The blue shaded region is excluded by our data at 95\% confidence level. The horizontal black line shows the limit from the CAST helioscope~\cite{Anastassopoulos2017}, and the grey line shows the ABRACADARBA limits on axion-like dark matter~\cite{Ouellet2019}. The trend of improved sensitivity to~$g_{a\gamma}$ at lower ALP masses (frequencies) is due to narrowing linewidth of the ALP signal, which scales as $\beta^2 \nu_a \approx 10^{-6} \nu_a$ in the standard halo model.
	}
	\label{fig:4}
\end{figure*}

The experimental run took place on September 12--17, 2019. After performing calibrations described above, we magnetized the four toroids with 6~A injected current in ($+ +$, $- -$) configuration and recorded 41 hours of data that were used to search for axion-like dark matter signals. Afterwards, we magnetized the four toroids with 6~A injected current in ($+ -$, $+ -$) configuration and collected 15 hours of data. Since in each channel the toroid magnetizations were now opposite, the axion-induced flux would vanish and we could use these data to confirm or reject any potential detection extracted from the first data set. We also collected 12 hours of data with no toroid magnetization, for the same purpose. In each data run, output voltages from the two SQUID detector channels were digitized at the sampling rate of 15.625~MS/s and recorded to a hard disk drive. We used the permeability sensing coils to monitor each toroid magnetization, and ensured that all magnetizations were unchanged before and after recording each data set. We note that experimental sensitivity to $g_{a\gamma}$ has a slow $1/4$-power scaling with averaging time, provided it exceeds ALP coherence time~\cite{Budker2014}.

Data analysis consisted of several processing, correction, and axion-like dark matter signal search steps. We divided the data into blocks of duration 2199~s ($2^{35}$ samples), chosen to be much longer than the ALP dark matter field coherence time for every frequency in the exclusion region. After removing 20 data blocks where SQUID feedback reset jumps occurred, we performed a discrete Fourier transform on each of the remaining 47 data blocks, used the SQUID detector transfer function $x$ to convert the frequency-domain data from voltage to magnetic flux $\Phi$, and formed symmetric and antisymmetric combinations of the two detector channels: $\Phi_{+}=(\Phi_{A}+\Phi_{B})/2$ and $\Phi_{-}=(\Phi_{A}-\Phi_{B})/2$. We calculated the power spectral density (PSD) for each of the four data streams $\Phi_{(A,B,+,-)}^2$, and averaged each PSD over all the blocks in the experimental run. We modeled each spectrum as the sum of an axion-like dark matter signal, SQUID detector noise, narrowband RF interference, and a broadband spectral baseline due to vibrational pickup of magnetic flux leakage from toroids. Our analysis assumed the boosted Maxwell-Boltzmann ALP signal lineshape predicted by the astrophysically-motivated standard halo model~\cite{Turner1990, Brubaker2017, Brubaker2017a, Du2018}. The linewidth of an axion-like dark matter signal at frequency~$\nu_a$ is determined by the parameter $\beta^2 \nu_a = 3v_0^2 \nu_a/(2c^2) \approx 10^{-6} \nu_a$, where $v_0 = 220~\text{km}/\text{s}$ is the local velocity of motion around the center of the galaxy. Following a simplified version of the data analysis approach used by the ADMX and HAYSTAC collaborations, we rejected narrowband RF interference and broadband spectral baseline by using Savitzky-Golay digital filtering to isolate these spectral features in the frequency domain~\cite{Brubaker2017a, Du2018}. By injecting simulated axion signals into the data, we determined that these corrections attenuated axion spectral signals by 7\%, and the final limits were adjusted accordingly~\cite{som}.

We performed the search for axion-like signals independently in each of the averaged spectra $\langle \Phi_{(A,B,+,-)}^2\rangle$ in the frequency range from 3~kHz to 3~MHz. In this range, each spectrum contained $6.6\times 10^9$ independent frequency points, which we broke up into 857 frequency bins, with bin width set to $10^4 \beta^2 \nu_b$, where $\nu_b$ is the bin center frequency. We modeled the histogram of PSD values in each bin as the chi-square distribution with 94 degrees of freedom, corresponding to the two quadratures of the Fourier transform averaged over 47 independent data blocks, fig.~\ref{fig:4}(a). We optimally filtered the data within each frequency bin by convolving with the standard halo model axion lineshape. We modeled the histogram of optimally-filtered data points as the normal distribution (verified by a Lilliefors test), and found its standard deviation $\sigma$ by fitting to the Gaussian dependence, fig.~\ref{fig:4}(b). We set the candidate detection threshold to $3.355\sigma$, equivalent to 95\% confidence interval for a $5\sigma$ detection~\cite{Brubaker2017a}, and flagged all points above the threshold as candidates. In the 3~kHz to 3~MHz frequency range, with our analysis parameters, there were 20043605 possible ALP frequency points. Assuming the optimally-filtered spectra are normally distributed, we would statistically expect 7954 candidate frequencies at which the spectral value is above the $3.355\sigma$ threshold in any given channel. Our analysis procedure produced 16292 candidates in channel A, 35375 candidates in channel B, 27278 candidates in the symmetric combination channel, and 12513 candidates in the antisymmetric combination channel. The extra candidates appeared due to residual electromagnetic interference and lower-frequency vibrational tones that got past our spectral correction procedure.

Resonant experiments, such as ADMX and HAYSTAC, perform re-scans of candidate frequencies to check if they are statistical deviations or a real axion signal. In our analysis, we used the two detector channels to make this decision. Axion-like dark matter detection claim would be made for candidates that are above threshold for the antisymmetric channel and in both channels A and B, but not in the symmetric channel. None of the candidates satisfied all four of these detection conditions. As a check of our data analysis procedure, we injected numerically-generated ALP signals into our data and verified that our analysis recovered the signals with their correct coupling strengths~\cite{som}. We used the conversion $\Phi_a = \omega_a g_{a\gamma} a_0 B_0 V / c$
between the ALP electromagnetic coupling $g_{a\gamma}$ and the induced magnetic flux $\Phi_a$ through the pickup coil, with the the mean magnetic field $B_0 = (1.51 \pm 0.03)~\text{T}$ at 6~A magnetizing current and the effective volume of each channel $V = (10.3 \pm 0.4)~\text{cm}^3$. We calculated the effective volume numerically, taking into account the $\approx 20\%$ suppression due to the Meissner effect in the lead superconducting magnetic shield~\cite{som}.

In the absence of a detection, we report for each frequency bin the magnitude of $g_{a\gamma}$ corresponding to the $5\sigma$ value of the optimally-filtered magnetic flux PSD in the antisymmetric channel, which represents the 95\% confidence interval limit for that bin, fig.~\ref{fig:4}(c). Before our work, the best experimental limit on the ALP-photon coupling for this mass range was set by the CERN CAST helioscope search for solar axions at $0.66\times 10^{-10}$~GeV$^{-1}$~\cite{Anastassopoulos2017}. Despite the much smaller volume of our apparatus (by a factor of $\approx 300$), our results constitute significant improvements on this limit over a broad range of ALP masses in the scenario that the ALP field is the dominant component of dark matter. Our experiment begins the exploration of the ALP mass and coupling region near $m_a \gtrsim 0.7~\text{neV}$ and $g_{a\gamma}\lesssim 8.8\times 10^{-10}~\text{GeV}^{-1}$, where existence of ALPs may resolve the tension between the observed TeV gamma-ray energy spectrum and the one expected based on the recent cosmic infrared background radiation data~\cite{Kohri2017}.

There are several ways to improve experimental sensitivity to axion-like dark matter. Cooling the SQUID magnetic sensors to milli-Kelvin temperature can reduce their noise by a factor of $\approx 10$, and it may be possible to achieve even better sensor performance using quantum upconversion~\cite{Chaudhuri2019b}. For an optimized search, it is necessary to use a scanned single-pole resonator to couple the ALP-induced magnetic flux to an amplifier operating at, or beyond, the Standard Quantum Limit~\cite{Chaudhuri2019}. The ferromagnetic toroidal core allowed us to achieve a 1.51~T static magnetic field with ampere-level injected current, which enabled the two-channel design, but magnetic material saturation prevented further field increase. New magnetic materials, with higher saturation field, could improve sensitivity further, or a high-field air-core toroidal superconducting magnet could be engineered, needing much higher injected current~\cite{Battesti2018}. The most dramatic sensitivity improvement could be achieved by increasing the toroid volume; the Fe-Ni magnetic alloy used in our experiment is inexpensive and readily available commercially. Scaling the volume up to the benchmark~$1~\text{m}^3$ would improve the sensitivity by another factor of~$10^4$. 

The authors acknowledge support from the NSF grant 1806557, the Heising-Simons Foundation grant 2015-039, the Simons Foundation grant 641332, and the Alfred P. Sloan foundation grant FG-2016-6728. The authors thank O.~P.~Sushkov for valuable comments on the manuscript.

\bibliography{library} 
 
\end{document}


\title{Supplementary Information for \\ ``Search for axion-like dark matter with ferromagnets''}

\author{Alexander~V.~Gramolin}
\affiliation{Department of Physics, Boston University, Boston, MA 02215, USA}
\author{Deniz~Aybas}
\affiliation{Department of Electrical and Computer Engineering, Boston University,Boston, MA 02215, USA}
\author{Dorian~Johnson}
\affiliation{Department of Physics, Boston University, Boston, MA 02215, USA}
\author{Janos~Adam}
\affiliation{Department of Physics, Boston University, Boston, MA 02215, USA}
\author{Alexander~O.~Sushkov}
\email{asu@bu.edu}
\affiliation{Department of Physics, Boston University, Boston, MA 02215, USA}
\affiliation{Department of Electrical and Computer Engineering, Boston University,Boston, MA 02215, USA}
\affiliation{Photonics Center, Boston University, Boston, MA 02215, USA}

\maketitle

\section{Macroscopic electromagnetism with an axion field}
\label{sec:electromagnetism}

Our experimental concept relies on a modification of Maxwell's equations by the axion-like dark matter field~$a$ coupled to the electromagnetic field~\cite{Sikivie1983,Wilczek1987}. In the presence of this coupling parametrized by $g_{a\gamma}$, the inhomogeneous Maxwell's equations take the form
\begin{align}
\begin{split}
\vec{\nabla}\cdot\vec{E} &= \frac{\rho_{\rm el}}{\epsilon_0} - cg_{a\gamma}\vec{B}\cdot\vec{\nabla}a, \\
\vec{\nabla}\times\vec{B}-\frac{1}{c^2}\pd{\vec{E}}{t} &= \mu_0\vec{j}_{\rm el} + \frac{g_{a\gamma}}{c}\left(\pd{a}{t}\vec{B} + \vec{\nabla}a\times\vec{E}\right),
\end{split}
\label{eq:211}
\end{align}
where $\rho_{\rm el}$ and $\vec{j}_{\rm el}$ are the usual electric charge and current densities associated with ordinary matter. We use SI units for electromagnetic fields and natural units for the axion field so that $g_{a\gamma}a$ is unitless, and the axion field amplitude $a_0$ is given by $m_a^2 a_0^2 / 2 = \rho_{\text{DM}} = 3.6 \times 10^{-42}~\text{GeV}^4$~\cite{PDG}. We neglect terms proportional to $\vec{\nabla}a$, since they are suppressed in the lumped-circuit case, and only consider the term proportional to $\partial a/\partial t$.

In the presence of electrically- and magnetically-polarizable materials, these equations should be written in terms of averaged charge and current densities. Following ref.~\cite{Jackson}, we average the microscopic electromagnetic fields that obey eqs.~(\ref{eq:211}) over a macroscopic sample, taking into account electric and magnetic multipole moments of the atoms inside the sample. As in standard electromagnetism, this averaging can be performed by introducing macroscopic material polarization $\vec{P}$ and magnetization $\vec{M}$, resulting in the averaged charge and current densities
\begin{align}
\begin{split}
\langle\rho_{\rm el}\rangle &= \rho_f - \vec{\nabla}\cdot\vec{P}+\ldots,\\
\langle\vec{j}_{\rm el}\rangle &= \vec{J}_f + \pd{\vec{P}}{t} + \vec{\nabla}\times\vec{M}+\ldots,
\end{split}
\end{align}
where $\rho_f$ and $\vec{J}_f$ are the macroscopic free charge and current densities, $\vec{\nabla}\cdot\vec{P}$ is the bound charge density, $\vec{\nabla}\times\vec{M}$ is the bound current density, and the ellipses indicate contributions from higher-order multipole moments. We can now define auxiliary fields $\vec{D} = \epsilon_0\vec{E}+\vec{P}$ and $\vec{H} = \vec{B}/\mu_0-\vec{M}$, resulting in Maxwell's equations in media, in the presence of the axion field:
\begin{align}
\begin{split}
\vec{\nabla}\cdot\vec{D} &= \rho_f,\\
\vec{\nabla}\times\vec{H}-\pd{\vec{D}}{t} &= \vec{J}_f + \frac{g_{a\gamma}}{\mu_0c}\frac{\partial a}{\partial t}\vec{B}.
\end{split}
\end{align}
In our work, we consider a material in which $\rho_f=0$ and focus on the second of the above equations, which now becomes
\begin{align}
\vec{\nabla}\times\vec{H}= \vec{J}_f+\frac{g_{a\gamma}}{\mu_0c}\frac{\partial a}{\partial t}\vec{B}.
\end{align}
This equation describes how, in the presence of a static magnetic field $\vec{B}$, the axion field $a = a_0 \sin{(\omega_a t)}$ is a source of a magnetic field oscillating at the same angular frequency~$\omega_a$~\cite{Sikivie1983}. This concept is behind many experimental searches for the axion-photon coupling $g_{a\gamma}$, and here we generalized it to experiments with magnetizable media. The key point is that inside a magnetic medium $\vec{B} = \mu_0(\vec{H}+\vec{M})$ includes the material magnetization, and in materials with large permeability this can be significantly enhanced compared to the free-space field $\mu_0\vec{H}$.

The boundary conditions for fields $\vec{B}$ and $\vec{H}$ at interfaces between different media are unchanged from the usual situation in electromagnetism (neglecting small effects of the axion field): the normal component of $\vec{B}$ is continuous, and the tangential component of $\vec{H}$ is continuous, unless there is a current density at the surface~\cite{Jackson}. Our toroidal geometry minimizes the effects of demagnetizing fields.

\section{Detailed description of the experimental apparatus}
\label{sec:apparatus}

Four ``High Flux 125$\mu$'' toroidal cores made of powdered Fe-Ni alloy were obtained from MH\&W International Corporation (part number CH778125E). We refer to these toroids as A1 (uppermost), A2, B1, and B2 (lowermost). The cores have a rectangular cross-section and the following dimensions: inner radius $r_1 = 24.4~\text{mm}$, outer radius $r_2 = 39.1~\text{mm}$, and height $h = 16.2~\text{mm}$. A magnetizing coil was wound around each toroid using 0.006" diameter polyimide-insulated Nb 48\% Ti wire obtained from California Fine Wire. Each magnetizing coil consisted of two counter-wound layers (to cancel out the azimuthal component of the current) with the following total number of turns: $N_{m}^{\text{A1}} = 1676$, $N_{m}^{\text{A2}} = 1678$, $N_{m}^{\text{B1}} = 1655$, and $N_{m}^{\text{B2}} = 1679$. In each detection channel, referred to as A (B), toroids A1 and A2 (B1 and B2) were stacked top-to-bottom. Within each channel, the two toroids could be independently magnetized in the same or the opposite directions by choosing polarities of currents in the magnetizing coils.

Nb-Ti twisted pairs and Nb washers were used to connect each magnetizing coil to a persistent switch placed within the outer superconducting shield. Each persistent switch consisted of a section of Nb-Ti wire encapsulated in Stycast 2850FT epoxy together with a 50~$\Omega$ surface-mount resistor (a persistent switch heater); applying a current pulse to the resistor made the Nb-Ti wire go from superconducting to normal and opened the persistent switch.

\begin{figure}[b!]
    \centering
    \includegraphics[width=0.85\textwidth]{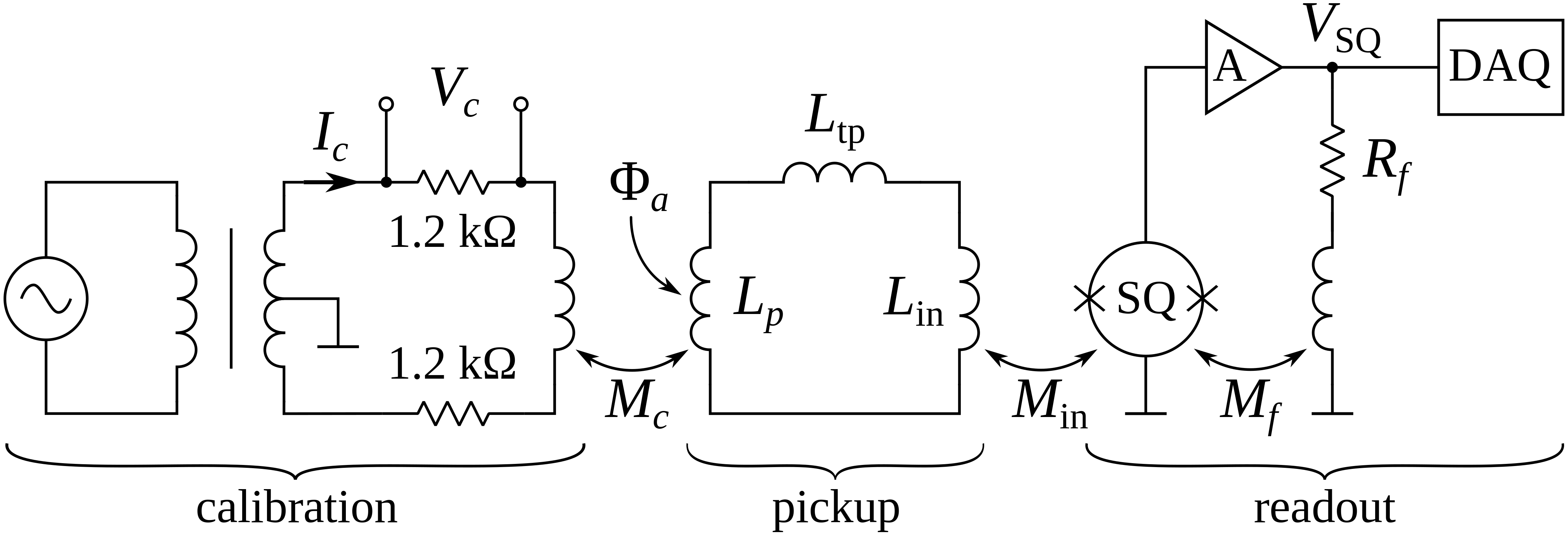}
    \caption{A schematic diagram of the experiment. The diagram is divided into three blocks representing the calibration circuit, the pickup coil, and the data readout system. Labels: $I_c$---current running through the calibration loop; $V_c$---voltage across a $1.2~\text{k}\Omega$ resistor, proportional to the calibration current ($V_c$ was monitored using the floating differential inputs of an SR560 voltage preamplifier); $\Phi_a$---axion-induced magnetic flux; SQ---SQUID; A---SQUID amplifier; $V_{\text{SQ}}$---SQUID output voltage; DAQ---data acquisition system; $L_p$---inductance of the pickup coil; $L_{\text{tp}}$---inductance of the twisted pair; $L_{\text{in}}$---inductance of the SQUID input coil; $R_f$---feedback resistance; $M_c$---mutual inductance between the calibration loop and the pickup coil; $M_{\text{in}}$---mutual inductance between the SQUID and its input coil; and $M_f$---mutual inductance between the SQUID and the feedback coil.}
    \label{fig:S1}
\end{figure}

The same Nb-Ti wire was used to wind two SQUID pickup coils on the G-10 support rod, on which the toroids were mounted using Stycast 2850FT epoxy. Each pickup coil was connected to a Magnicon single-stage SQUID sensor using a twisted pair, fig.~\ref{fig:S1}. The parameters of the pickup coils and the SQUID sensors are summarized in table~\ref{tab:S1}. During data collection, the SQUIDs were operated in a flux-locked-loop (FLL) mode with the amplifier feedback resistance $R_f = 30~\text{k}\Omega$ and the gain-bandwidth product $7.20~\text{GHz}$. Experimental calibration of the SQUID frequency response is described in section~\ref{sec:squidcal}.

\begin{table}
	\centering
	\setlength{\tabcolsep}{5pt}
	\caption{Parameters of the pickup coils and the SQUID sensors.}
	\label{tab:S1}
	\begin{tabular}{cccccccc}
		\hline\hline
		Detection & $N_p$ & $L_p$ & $L_{\text{tp}}$ & SQUID & $L_{\text{in}}$ & $M_{\text{in}}$ & $M_f$ \\
		channel & & (nH) & (nH) & serial~\# & (nH) & (nH) & (nH) \\ \hline
        A & 6 & 3320 & 100 & S0142 & 1800 & 8.54 & 0.047 \\
        B & 6 & 3320 & 100 & S0141 & 1800 & 8.65 & 0.046 \\
		\hline\hline
	\end{tabular}
\end{table}

During data collection runs, we used a custom cryogenic switch to disconnect all the twisted pair leads (other than those of the pickup coils) entering the inner superconducting shield. This allowed us to reduce the RF interference signals entering the experiment. The cryogenic switch, placed within the outer superconducting shield, consisted of 50 individual reed switches manufactured by Standex Electronics (part number KSK-1A85-2030). These reed switches were placed inside a G-10 tube, which was filled with Stycast 2850FT epoxy. On the outer surface of the tube, a solenoidal coil was wound using copper magnet wire. By running a 2.5~A current through this coil, we could activate the reed switches, which were normally open. For the magnetizing coils, we used three reed switches connected in parallel (six switches in total per a twisted pair), which allowed us to achieve the carry current of 10~A at liquid-helium temperature.

When the toroids were magnetized, some fraction of the magnetic field leaked out and coupled into the pickup coils. In order to address the issue of seismic vibrations causing the pickup coils to move relative to this stray magnetic field, a tripod was constructed to support the cryostat housing the apparatus by connecting an aluminum frame to three Newport I-2000 high performance laminar flow isolators. Seismic isolation of the apparatus from the ground led to reduction of vibrational magnetic pickup at acoustic frequencies, which simplified the procedure of locking SQUID electronics feedback loops.

\section{Measurements of toroid permeability}
\label{sec:permeability}

\begin{figure}[b!]
	\centering
	\includegraphics[width = 0.7\textwidth]{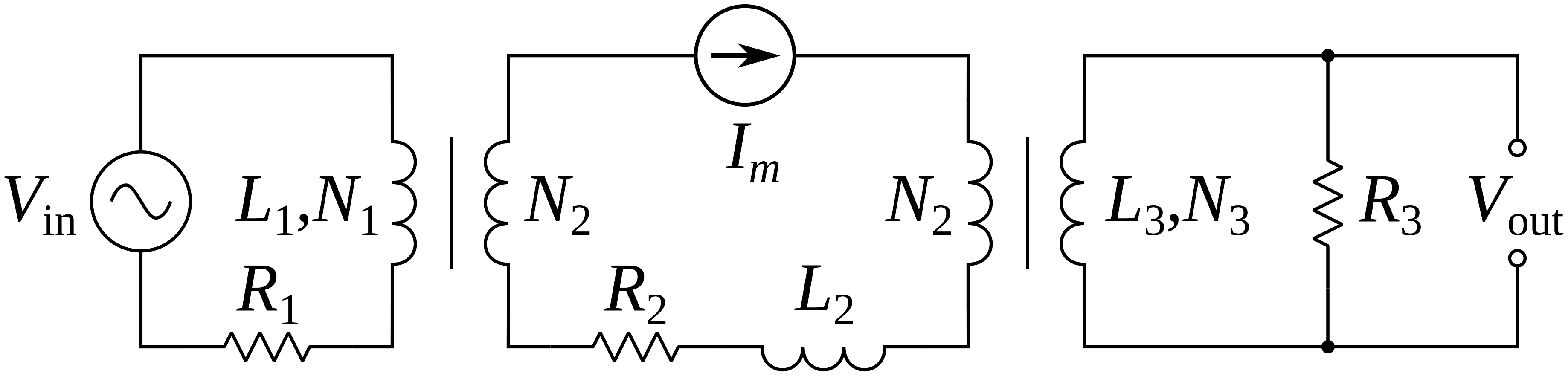}
	\caption{\label{fig:S2}A schematic diagram of the circuit used to measure inductances of the toroid magnetizing coils while injecting a DC magnetizing current~$I_m$.}
\end{figure}

The magnetic permeability of the core material can be extracted from the self-inductance of the magnetizing coil. To measure the latter, we used the MFLI lock-in amplifier from Zurich Instruments connected as shown in fig.~\ref{fig:S2}. The DC current $I_m$ ranging from 0 to 10~A was injected into the superconducting magnetizing coil~$L_2$ using a model KA3010D power supply from Korad Technology. The resistance $R_2$ in fig.~\ref{fig:S2} accounts for the internal impedance of the power supply, the resistance of current leads, and transformer losses. The reference AC signal $V_{\text{in}}$ from the lock-in amplifier was coupled to the circuit through a toroidal-core transformer with the winding turn ratio $N_1 : N_2 = 40 : 3$ and the primary coil inductance~$L_1 = 258~\mu\text{H}$. The output voltage~$V_{\text{out}}$ was measured by the lock-in amplifier using a transformer with $N_2 : N_3 = 3 : 214$ and $L_3 = 7.56~\text{mH}$. Both transformers were wound on high saturation flux density iron-powder toroidal cores (T200A-26 obtained from Amidon Corp.). The resistances $R_1 = 50~\Omega$ and $R_3 = 10~\text{M}\Omega$ were the corresponding internal impedances of the lock-in amplifier.

\begin{figure}[t!]
	\centering
	\includegraphics[width=0.49\textwidth]{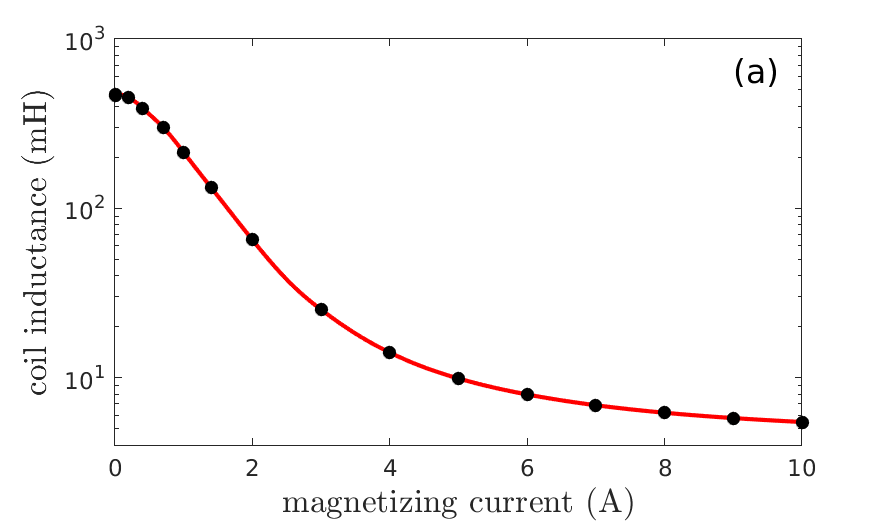}
	\includegraphics[width=0.49\textwidth]{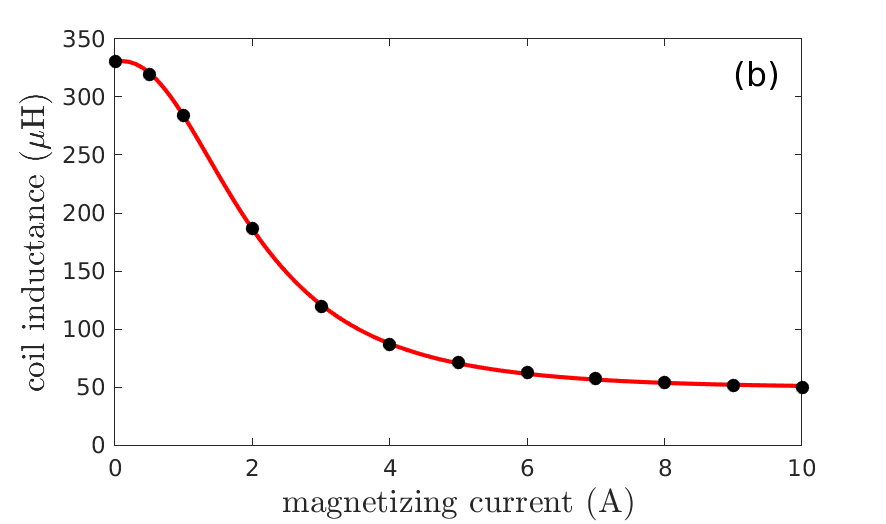}
	\caption{(a) Measurements of the toroid A1 magnetizing coil inductance as a function of injected current. (b) Measurements of the toroid A1 permeability sensing coil inductance as a function of current injected into the magnetizing coil.}
	\label{fig:S3}
\end{figure}

By solving the circuit shown in fig.~\ref{fig:S2}, we obtain the following expression for the output voltage $V_{\text{out}}$:
\begin{equation}
V_{\text{out}} \approx - \frac{N_2^2}{N_1 N_3} \frac{\omega^2 L_1 L_3 R_3}{(R_1 + i\omega L_1) (R_2 + i\omega L_2) (R_3 + i\omega L_3)} V_{\text{in}}, \label{Eq_Vout}
\end{equation}
where $\omega$ is the angular frequency of the input signal $V_{\text{in}}$. Both the real (in-phase) and imaginary (out-of-phase) components of $V_{\text{out}}$ were measured using the lock-in amplifier in the frequency range of 100~Hz to 10~kHz. These data were fitted with the corresponding functions following from eq.~(\ref{Eq_Vout}), where $R_2$ and $L_2$ were the fit parameters. The resulting best-fit inductance values of the coil are shown in fig.~\ref{fig:S3}(a) for different values of the magnetizing current~$I_m$.

The effective magnetic permeability of the core can be calculated using a well-known formula for the self-inductance of a toroidal coil with a rectangular cross-section as 
\begin{equation}
\mu = \frac{2\pi L_2}{\mu_0 N_m^2 h \log{(r_2 / r_1)}}.
\end{equation}
The corresponding results for $\mu$ are shown in fig.~2(a) of the main text.

During data collection, we used the permeability sensing coils wound on each toroid to monitor their magnetization. Figure~\ref{fig:S3}(b) shows the self-inductance of the toroid A1 permeability sensing coil, measured at different magnetizing currents using the lock-in amplifier. Each permeability sensing coil contained 30 turns wound clockwise and another 30 turns placed diagonally across the toroid and wound counter-clockwise, which allowed us to reduce coupling between the sensing and magnetizing coils. Without this precaution, the inductance of the sensing coil would be greatly suppressed when the magnetizing coil is in a persistent current mode (this can be thought as a transformer whose secondary coil is shorted).

\section{Axion-induced magnetic flux through the pickup coil}
\label{sec:flux}

The effect of the axion-like dark matter field on the measurement apparatus is identical to that of an effective azimuthal current density
\begin{equation}
J_{\text{eff}} = \frac{g_{a\gamma}}{\mu_0 c} \, \frac{\partial a}{\partial t} B_T = \frac{\omega_a}{\mu_0 c} g_{a\gamma} a_0 \cos{(\omega_a t) B_T}, \label{Eq.Jeff}
\end{equation}
where $B_T$ is the azimuthal magnetic field inside the toroidal sample. Measurements of this field for the magnetizing coil current~$I_m$ up to 10~A are shown in fig.~2(b) of the main text. We note that the auxiliary field $H$ varies across the toroid cross-section: 
\begin{align}
H(s) = \frac{N_mI_m}{2\pi s},
\end{align}
where $s$ is the distance to the toroid axis; $r_1 < s < r_2$ inside the toroid. Therefore, $B_T$ also depends on~$s$:
\begin{align}
B_T(s) = \mu_0 \left[H(s) + M_0 \tanh{\frac{H(s)}{H_0}}\right],
\end{align}
with parameters $M_0 = (1.15 \pm 0.03) \times 10^6~\text{A}/\text{m}$ and $H_0 = (1.20 \pm 0.05) \times 10^4~\text{A}/\text{m}$, as given in the main text. Since at the magnetizing coil current of 6~A the permeable material is close to saturation, the variation of~$B_T$ across the toroid cross-section is small: $B_T = 1.53~\text{T}$ at the inner radius and $B_T = 1.49~\text{T}$ at the outer radius.

The effective current $J_{\text{eff}}$ creates the magnetic field~$B_a$ oscillating at the frequency $\nu_a = \omega_a / (2\pi)$ and oriented along the axis of the toroidal sample. The field $B_a$ can be calculated using the following expression for a magnetic field of a circular current loop lying in the $xy$~plane with its center at the origin:
\begin{equation}
B_z (r, \, z) = \frac{\mu_0 I}{4\pi} \int\limits_{0}^{2\pi} \frac{s (s - r \cos{\theta})}{(z^2 + r^2 + s^2 - 2r s \cos{\theta})^{3/2}} \, d\theta, 
\end{equation}
where $r = \sqrt{x^2 + y^2}$, $I$ is the current running through the loop, and $s$~is its radius. Then in the mid-plane of the sample,
\begin{align}
B_a (r) = \frac{\mu_0}{4\pi} \int\limits_{r_1}^{r_2} ds \, J_{\text{eff}} \int\limits_{-h}^{h} dz \int\limits_{0}^{2\pi} d\theta \, \frac{s (s - r \cos{\theta})}{(z^2 + r^2 + s^2 - 2r s \cos{\theta})^{3/2}} 
= \frac{\mu_0 h}{2\pi} \int\limits_{r_1}^{r_2} ds \int\limits_{0}^{2\pi} d\theta \, J_{\text{eff}}\,\frac{s (s - r \cos{\theta})}{{\tilde r}^2 \sqrt{h^2 + {\tilde r}^2}},
\end{align}
where ${\tilde r}^2 = r^2 + s^2 - 2r s \cos{\theta}$.
The oscillating axion-induced magnetic flux through the pickup loop of radius~$r_p$ is
\begin{equation}
\Phi_a = \int\limits_{0}^{r_p} dr \, 2\pi r B_a(r) = \frac{\omega_a}{c} g_{a\gamma} a_0 \cos(\omega_a t) h \int\limits_{0}^{r_p}dr\int\limits_{r_1}^{r_2} ds \int\limits_{0}^{2\pi} d\theta \, B_T(s)\, \frac{ rs (s - r \cos{\theta})}{{\tilde r}^2 \sqrt{h^2 + {\tilde r}^2}}.
\label{eq:567}
\end{equation}
The flux~$\Phi_a$ depends on the toroid and pickup loop geometry, as well as on the magnetic saturation properties of the toroid core permeable material, which affect~$B_T(s)$. We parametrize these factors in terms of the effective volume~$V$, defined as
\begin{align}
V = h \int\limits_{0}^{r_p}dr\int\limits_{r_1}^{r_2} ds \int\limits_{0}^{2\pi} d\theta \, \frac{B_T(s)}{B_0} \frac{ rs (s - r \cos{\theta})}{{\tilde r}^2 \sqrt{h^2 + {\tilde r}^2}}, \label{Eq_V}
\end{align}
where $B_0 = B_T((r_1+r_2)/2)$ is the magnetic field at the center of the toroid cross-section. With this definition of the effective volume and magnetic field inside the toroid, the flux induced by the axion-like dark matter is given by
\begin{align}
\Phi_a = \frac{\omega_a}{c} g_{a\gamma} a_0 \cos(\omega_a t) B_0 V. \label{eq:Phi_a}
\end{align}

For the sample parameters
\begin{equation}
\begin{gathered}
r_1 = 24.4 \pm 0.2~\text{mm}, \quad r_2 = 39.1 \pm 0.2~\text{mm}, \quad h = 16.2 \pm 0.2~\text{mm}, \quad r_p = 23.5 \pm 0.1~\text{mm}, \\
\quad I_m = 6 \pm 0.01~\text{A}, \quad N_m = \frac{1}{4} \left(N_{m}^{\text{A1}} + N_{m}^{\text{A2}} + N_{m}^{\text{B1}} + N_{m}^{\text{B2}}\right) = 1672 \pm 12,
\end{gathered}
\end{equation}
the numerical value of the effective volume calculated using eq.~(\ref{Eq_V}) is $V = (13.5 \pm 0.3)~\text{cm}^3$.  At the 6~A magnetizing coil current, the magnetic field at the center of the toroid cross-section is $B_0 = (1.51 \pm 0.03)~\text{T}$.

If the experiment were performed with an air-core toroidal magnet of the same dimensions and the same current, then the effective volume would be $V^{\text{(air)}} = 13.9 \pm 0.3~\text{cm}^3$ and the magnetic field at the center of the toroid cross-section would be $B_0^{\text{(air)}} = 0.063~\text{T}$. The physical volume of each sample is $2\pi h (r_2^2 - r_1^2) = 95~\text{cm}^3$.

\section{Calculation of inductances and effective volume using COMSOL}
\label{sec:comsol}

\begin{figure}
    \centering
    \includegraphics[width=0.85\textwidth]{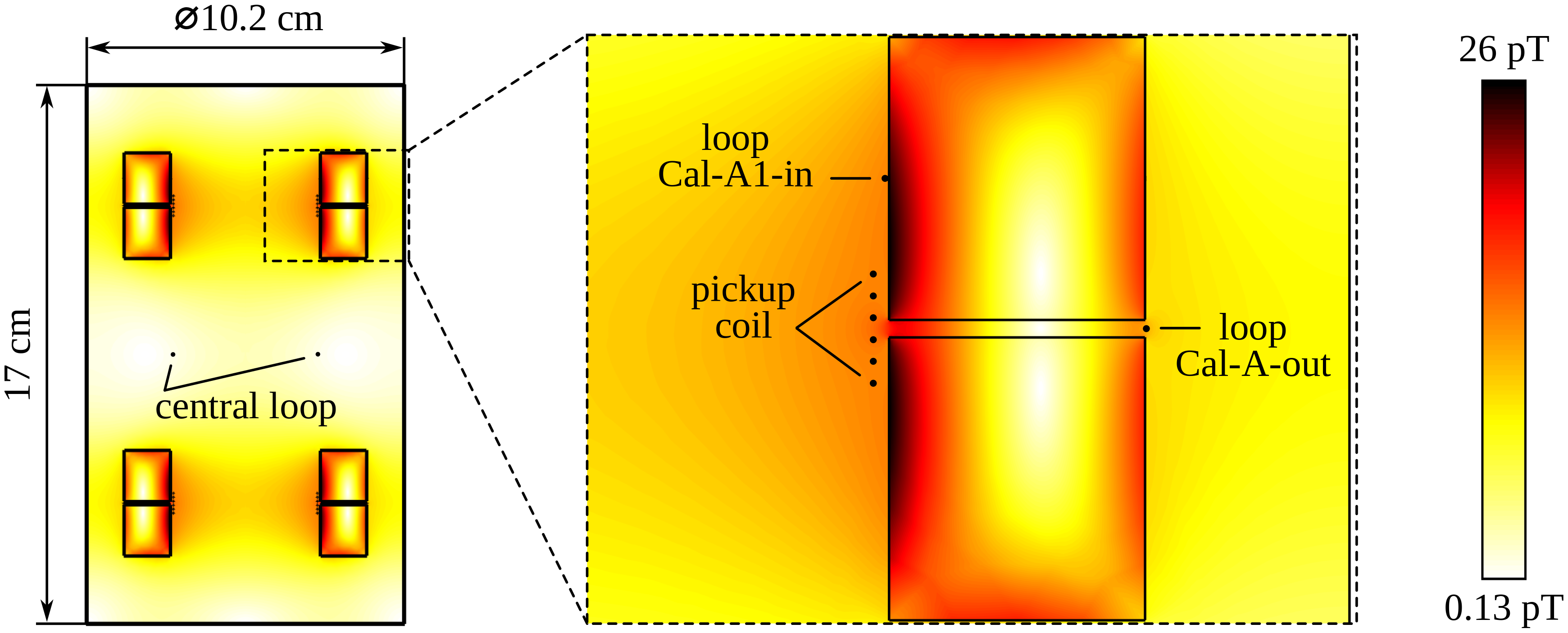}
    \caption{Results of a COMSOL simulation of the magnetic field created by the effective current density $J_{\text{eff}} = 1.6~\text{nA}/\text{m}^2$ running through the toroids, corresponding to an axion-like dark matter field with the coupling strength $g_{a\gamma}=10^{-10}$~GeV$^{-1}$ at the Compton frequency $\nu_a = 10~\text{kHz}$. A vertical section of the cylindrically-symmetric geometry is shown. The magnitude of the magnetic field is indicated by the color scale shown on the right. Also shown are the locations of the pickup coil and the calibration loops.}
    \label{fig:S4}
\end{figure}

COMSOL Multiphysics software was used to perform a numerical simulation of the relevant parameters for our experimental geometry, fig.~\ref{fig:S4}. All simulations used the ``physics-controlled'' mesh type and the ``normal'' element size. The inner superconducting shield was emulated by a material with a nearly zero magnetic permeability, $\mu = 10^{-10}$. Magnetic properties of the toroids were taken into account by using a material with $\mu = 1.9$ (this corresponds to the magnetizing current of 6~A, see fig.~2(a) of the main text).

Table~\ref{tab:S2} shows results of calculations of the mutual inductances $M_c$ between different calibration loops and the corresponding pickup coils. These values were necessary for calibrating the SQUID flux-to-voltage transfer function~$x$, as described in section~\ref{sec:squidcal}. Uncertainties in the simulated values of $M_c$ were estimated by varying the apparatus dimensions and the toroid magnetization. The dominant contribution is due to the uncertainty in the radius of the inner superconducting shield.

\begin{table}[b!]
	\centering
	\setlength{\tabcolsep}{5pt}
	\caption{Mutual inductances between the calibration loops and the pickup coils.}
	\label{tab:S2}
	\begin{tabular}{ccccc}
		\hline\hline
		& Central & Loop & Loop & Loop \\
		& loop & Cal-A1-in & Cal-A-out & Cal-B-out \\
		\hline
		$M_c$ (nH) & $7.4 \pm 0.3$ & $203 \pm 2$ & $104 \pm 4$ & $104 \pm 4$ \\
		\hline\hline
	\end{tabular}
\end{table}

The influence of the superconducting shield made it important to carry out numerical calculations of the effective volume. The simulated magnetic field profile is shown in fig.~\ref{fig:S4} for an effective azimuthal current density $J_{\text{eff}} = 1.6~\text{nA}/\text{m}^2$ inside each toroid, corresponding to an axion-like dark matter field with the coupling strength $g_{a\gamma}=10^{-10}$~GeV$^{-1}$ at the Compton frequency $\nu_a = 10~\text{kHz}$. The flux $\Phi_a$ was obtained by integrating~$B_a$ over the area of the pickup coil, and the effective volume was calculated using eqs.~(\ref{Eq.Jeff}) and (\ref{eq:Phi_a}) as $V = \Phi_a / (\mu_0 J_{\text{eff}})$. The results are shown in table~\ref{tab:S3}. We considered two cases: Fe-Ni toroids magnetized with a 6~A current (our experimental situation) and hypothetical air-core toroidal coils with the same dimensions and current. In each case, we performed two simulations, with and without the inner superconducting shield. The no-shield values of the effective volume are in agreement with the analytical calculations presented in section~\ref{sec:flux}. The presence of the inner superconducting shield reduced the flux and the values of the effective volume by $\approx 20\%$. The uncertainty of~4\% in the simulated values of~$V$ was estimated by varying the apparatus geometry, the radial distribution of the effective current, and the toroid magnetization.

All our calculations of the axion-like dark matter signal magnitude used the effective volume of $(10.3 \pm 0.4)~\text{cm}^3$, taking into account the presence of the superconducting shield.

\begin{table}[h]
	\centering
	\setlength{\tabcolsep}{5pt}
	\caption{Effective volumes $V$ (in $\text{cm}^3$) calculated using COMSOL.}
	\label{tab:S3}
	\begin{tabular}{ccc}
		\hline\hline
		& \multirow{2}{*}{Air core} & Fe-Ni core, \\
		& & $I_m = 6~\text{A}$ \\
		\hline
		No superconducting shield & $13.1 \pm 0.5$ & $12.9 \pm 0.5$ \\
		With the superconducting shield & $10.4 \pm 0.4$ & $10.3 \pm 0.4$ \\
		\hline\hline
	\end{tabular}
\end{table}

\section{Calibration of SQUID flux-to-voltage transfer function}
\label{sec:squidcal}

In our experiments, the SQUIDs were used as magnetic flux-to-voltage converters, characterized by a transfer function~$x$ that converts the magnetic flux~$\Phi$ through the pickup coil to the output voltage~$V_{\text{SQ}}$ of the SQUID feedback circuit:
\begin{equation}
V_{\text{SQ}} = x \, \Phi = x \iint\limits_{\substack{\text{pickup} \\ \text{coil area}}} B \, dS,
\end{equation}
where $B$ is the axial magnetic field (fig.~1 of the main text). The frequency-dependent function $x(\nu)$ depends on the coupling of the pickup coil to the SQUID and on the feedback circuit parameters. Direct measurements of $x(\nu)$ were performed by injecting a sinusoidally-varying current~$I_c$ of known amplitude and frequency into one of the calibration loops and recording the SQUID output voltage. We used a center-tapped transformer to provide DC isolation between the signal generator and the calibration loop, fig.~\ref{fig:S1}. The calibration procedure was performed for all four loops at frequencies up to 4~MHz. At each frequency, the transfer coefficient~$x$ was calculated using the relation
\begin{equation}
V_{\text{SQ}} = x \frac{M_c I_c}{N_p},
\end{equation}
where $N_p = 6$ is the number of turns of the pickup coil and $M_c$~is the mutual inductance between the calibration loop and the pickup coil, fig.~\ref{fig:S1}. The procedure for calculating mutual inductances is described in section~\ref{sec:comsol}; for each calibration loop, the value of $M_c$ was used as listed in table~\ref{tab:S2}.

The frequency dependence of the transfer function $x(\nu)$ was modeled as a 2-pole low-pass filter:
\begin{align}
    x(\nu) = \frac{x_0}{(1-\nu^2/\nu_0^2)+i\nu/(Qb)}, \label{eq:x_nu}
\end{align}
where $x_0$ is the low-frequency value of the transfer function, $\nu_0$ is the natural frequency, and $Q$ is the quality factor. The real part of $x$ corresponds to the in-phase quadrature, and the imaginary part corresponds to the out-of-phase quadrature of the SQUID response, relative to the applied signal used as a reference. The best-fit parameters and the fit uncertainties are listed in table~\ref{tab:S4} for the central loop calibration data. The frequency dependence of $|x(\nu)|$ is plotted in fig.~3(b) of the main text. 

\begin{table}[t!]
	\centering
	\setlength{\tabcolsep}{5pt}
	\caption{Parameters for $x(\nu)$ obtained using the central calibration loop.}
	\label{tab:S4}
	\begin{tabular}{cccc}
		\hline\hline
		Detection & $x_0$ & $\nu_0$ & $Q$ \\
		channel & ($\times 10^{12}~\text{V/Wb}$) & ($\times 10^{6}~\text{Hz}$) & \\
		\hline
		A & $5.13 \pm 0.15$ & $1.98 \pm 0.04$ & $0.52 \pm 0.02$ \\
		B & $5.94 \pm 0.06$ & $1.92 \pm 0.03$ & $0.61 \pm 0.03$ \\
		\hline\hline
	\end{tabular}
\end{table}

There was $\approx 25\%$ variation in the transfer function magnitude extracted using the different calibration loops, see table~\ref{tab:S5}. The greatest deviation was measured for the ``Cal-A1-in'' loop, which was placed at the inner circumference of toroid A1, under its superconducting magnetizing coil. This placement was chosen to evaluate the distortion and screening of magnetic fields by the superconducting magnetizing coil. For our final transfer function parameters, we chose the conservative values listed in table~\ref{tab:S4} and assigned a 25\% uncertainty to~$x_0$. We note that if we had used the value $x_0 = 7.0 \times 10^{12}~\text{V}/\text{Wb}$ then we would have obtained a stronger limit on $g_{a\gamma}$.

\begin{table}[b!]
	\centering
	\setlength{\tabcolsep}{5pt}
	\caption{SQUID transfer function magnitude for different calibration loops.}
	\label{tab:S5}
	\begin{tabular}{ccc}
		\hline\hline
		Detection & Calibration & $x_0$ \\
		channel & loop & ($\times 10^{12}~\text{V/Wb}$) \\
		\hline
		A & Central & 5.1 \\
		A & Cal-A1-in & 7.0 \\
		A & Cal-A-out & 5.8 \\
		B & Central & 5.9 \\
		B & Cal-B-out & 5.9 \\
		\hline\hline
	\end{tabular}
\end{table}

The high-frequency roll-off of the transfer function is due to the finite bandwidth of the SQUID feedback electronics. The low-frequency magnitude of~$x$ can be estimated by considering the circuit that couples the magnetic flux through the pickup coil into the SQUID, fig.~\ref{fig:S1}. The flux $\Phi_{\text{SQ}}$ through the SQUID is proportional to the flux $\Phi$ through the pickup coil:
\begin{equation}
\Phi_{\text{SQ}} = \frac{N_p M_{\text{in}}}{L_p + L_{\text{tp}} + L_{\text{in}}} \Phi,
\end{equation}
where $N_p=6$ is the number of turns in the pickup coil, $M_{\text{in}}$~is the mutual inductance between the input coil and the SQUID, $L_p$~is the pickup coil self-inductance, $L_{\text{tp}}$~is the inductance of the twisted pair that connects the pickup coil to the SQUID input coil, and $L_{\text{in}}$~is the input coil self-inductance. In the flux-locked-loop (FLL) mode, the flux through the SQUID is proportional to the feedback voltage:
\begin{equation}
\Phi_{\text{SQ}} = \frac{M_f}{R_f} V_{\text{SQ}},
\end{equation}
where $M_f$~is the mutual inductance between the feedback coil and the SQUID, and $R_f$~is the feedback resistance. Therefore, the low-frequency value of the SQUID transfer function can be estimated as
\begin{equation}
x_0 = \frac{R_f}{M_f} \frac{N_p M_{\text{in}}}{L_p + L_{\text{tp}} + L_{\text{in}}} \approx 6.3 \times 10^{12}~\text{V/Wb}, \label{eq:x_estimate}
\end{equation}
where $R_f = 30~\text{k}\Omega$ and the other parameters are taken from table~\ref{tab:S1}. This estimate is consistent with the measured values quoted in table~\ref{tab:S5}. 

Note that, since $L_p$ in the denominator of eq.~(\ref{eq:x_estimate}) is roughly proportional to~$N_p^2$, $x_0$ has a non-linear dependence on the number of pickup coil turns, $N_p$. During experimental design, a numerical optimization of $N_p$ was performed, and the chosen value $N_p = 6$ maximized the magnitude  of~$x_0$. For this coupling circuit design, the flux coupling from pickup coil to SQUID is: $\Phi_{\text{SQ}}\approx\Phi/100$.

\section{Expected axion-like dark matter signal lineshape}

In order to obtain the characteristics of the expected experimental signal, laboratory searches for the axion-like dark matter halo make assumptions about the local dark matter velocity and mass distributions. There are several models for these distributions. In our work, we assume the standard halo model: a fully virialized halo that is spherically symmetric and approximately isothermal, so that in the galactic reference frame dark matter velocities obey the Maxwell-Boltzmann distribution, neglecting the cutoff at the galactic escape velocity~\cite{Jimenez2003}. The shape of this distribution is specified by the local circular velocity $v_0=220$~km/s and the local dark matter density $\rho_{\rm DM}=0.45$~GeV/cm$^3 = 3.6\times10^{-42}$~GeV$^4$~\cite{PDG}. The motion of the terrestrial laboratory frame with respect to the galactic rest frame is dominated by the velocity~$v_0$ of the Sun rotating around the center of the galaxy. In the terrestrial frame, the dark matter velocity distribution is not, in general, Maxwellian. The spectral shape of the expected axion-like dark matter signal in the laboratory frame is
\begin{align}
    f(\nu) = \frac{1}{\sqrt{\pi}}\left( \sqrt{\frac{3}{2}}\frac{1}{r\nu_a\beta^2}\right)
    \sinh\left( 3r\sqrt{\frac{2(\nu-\nu_a)}{\nu_a\beta^2}}\right)
    \exp{\left( -\frac{3(\nu-\nu_a)}{\nu_a\beta^2}-\frac{3r^2}{2}\right)},
\label{eq:100}
\end{align}
where $\nu_a=m_ac^2/h$ is the Compton frequency of the axion-like field, $r \approx \sqrt{2/3}$, and $\beta^2 = 3v_0^2 / (2c^2) \approx 0.8 \times 10^{-6}$ is the parameter that sets the spectral linewidth of the axion signal~\cite{Turner1990, Brubaker2017, Brubaker2017a}. The lineshape function used for optimal filter convolution was normalized to the value of $1/2$, to satisfy Parseval's theorem.

\section{Data analysis}

During data taking, the output voltage of each Magnicon SQUID was digitized with a Spectrum Instrumentation M4i.4421-x8 ADC card, sampling at 15.625~MS/s, and stored to a hard disk drive. Data analysis calculations were performed using MATLAB on the Shared Computing Cluster, which is administered by Boston University’s Research Computing Services. The data processing procedure consisted of the following steps.
\begin{enumerate}[(1),noitemsep,leftmargin=*] 
	\item
    Divide the 41 hours of data into 67 blocks. Each block duration was 2199~s ($2^{35}$ samples), chosen to be much longer than the ALP dark matter field coherence time for every frequency in the exclusion region: at the lower boundary of 3~kHz there are 7 coherence times per block, and at higher ALP Compton frequencies this ratio is larger.
    \item
    Discard the data blocks where a SQUID reset jump occurs in either of the two SQUID channels (20 blocks were discarded, 47 blocks remained, corresponding to 29 hours of data). These reset jumps occurred when the SQUID feedback output voltage exceeded the $\pm 10$~V dynamic range of the feedback electronics. 
    \item
    Perform a discrete Fourier transform on each data block to transform the voltage data, $V(t)$, into the frequency domain, $V(\nu)$. No windowing was applied. 
    \item
    For each data block and detector channel, use the SQUID detector transfer function to convert the frequency-domain data from voltage to magnetic flux: $\Phi(\nu) = V(\nu)/x(\nu)$.
    \item
    Form a symmetric and an antisymmetric combinations of the two detector channels, $\Phi_{+}=(\Phi_{A}+\Phi_{B})/2$ and $\Phi_{-}=(\Phi_{A}-\Phi_{B})/2$, respectively. The antisymmetric combination preserves the ALP dark matter signal and suppresses common-mode RF interference, whereas the symmetric combination suppresses the ALP dark matter signal.
    \item
    Form the power spectral density (PSD) for each of the four data streams: $\Phi_{A}^2$, $\Phi_{B}^2$, $\Phi_{+}^2$, and~$\Phi_{-}^2$. Average the PSDs over the 47 data blocks in the experimental run to form an average spectrum for each data stream: $\langle \Phi_{A}^2 \rangle$, $\langle \Phi_{B}^2 \rangle$, $\langle \Phi_{+}^2 \rangle$, and~$\langle \Phi_{-}^2 \rangle$.
\end{enumerate}

Each spectrum was modeled as a sum of four contributions: $\langle \Phi^2\rangle = \Phi_n^2 + \Phi_i^2 + \Phi_b^2 + \Phi_a^2$, where $\Phi_n^2$ is the SQUID magnetometer noise PSD, $\Phi_i^2$ is the narrowband RF interference, $\Phi_b^2$ is a broadband spectral baseline due to vibrational pickup of the magnetic flux leakage from the toroids, and $\Phi_a^2$~is the expected ALP signal following the spectral shape of eq.~(\ref{eq:100}). The fact that the standard halo model ALP signal linewidth is $\approx \beta^2\nu_a$ was used to perform two spectral corrections: a rejection of the narrowband RF interference~$\Phi_i^2$ with a spectral feature width~$\lesssim 0.1\beta^2\nu_a$ and a subtraction of the broadband spectral baseline~$\Phi_b^2$ with a spectral feature width~$\gtrsim 30\beta^2\nu_a$. Savitzky-Golay digital filtering was used to differentiate between these spectral features. This correction procedure, described below, was a simplified version of the analysis approach used by the ADMX and HAYSTAC collaborations~\cite{Brubaker2017, Brubaker2017a, Du2018}.

The search for an axion signal was performed independently in each of the averaged spectra $\langle \Phi_{(A,B,+,-)}^2\rangle$ between 3~kHz and 3~MHz. In this range, each spectrum contained $6.6\times 10^9$ independent frequency points. The entire range was broken up into 857 frequency bins, with the bin width set to $10^4 \beta^2 \nu_b$, where $\nu_b$~is the bin center frequency. Since the linewidth of the expected axion signal in the standard halo model~(\ref{eq:100}) is $\approx \beta^2 \nu_a \approx 10^{-6} \nu_a$, each bin contains $\approx 10^4$ distinct axion search frequencies, and the fractional variation of the ALP search frequency over one bin is~$\approx 1\%$.

A histogram of the PSD values in each bin was modeled as the chi-square distribution with 94 degrees of freedom, fig.~4(a) of the main text. This distribution is expected because the PSD is formed by adding the squares of the real and imaginary parts of the Fourier transform and averaging over 47 data blocks, therefore, it is a sum of squares of 94 independent normally-distributed random variables~\cite{Walck2007}. The following two spectral correction steps were performed for each frequency bin.
\begin{enumerate}[(1),noitemsep,leftmargin=*] 
	\item
    Search for narrow peaks using a Savitzky-Golay filter with the order~4 and the impulse response half-length $0.16\beta^2\nu_b$. These filter parameters were chosen so that the 3~dB cutoff frequency of the filter was $10\times$ the expected axion signal linewidth $\beta^2\nu_b$~\cite{Schafer2011}. Therefore, once the filtered data were subtracted, only spectral features narrower than $0.1\beta^2\nu_b$ remain. After subtraction, the distribution of the PSD values was fitted to a shifted chi-squared distribution. The PSD values were designated as the narrowband RF interference peaks and removed if their magnitude was above the cutoff threshold, set at the chi-squared distribution value of~0.1. This procedure was repeated iteratively, since large narrow peaks can lead to large local spectral baseline distortions, but it normally converged after 1-3 iterations. This correction removed 0.02\% of the data points. By injecting simulated axion signals into the data, it was determined that this correction does not affect axion spectral signals.
    \item
    Fit and subtract a broad spectral baseline, which was determined by filtering the data with a Savitzky-Golay filter with the order~4 and the impulse response half-length~$47 \beta^2 \nu_b$. These filter parameters were chosen so that the 3~dB cutoff frequency of the filter was $0.03\times$ the expected axion signal linewidth $\beta^2\nu_b$~\cite{Schafer2011}. Therefore, the baseline subtraction preserved spectral features narrower than $30\beta^2\nu_b$, with some attenuation. By injecting simulated axion signals into the data, it was determined that this broad baseline subtraction in fact attenuated axion spectral signals by 7\%, and the final limits were adjusted accordingly, see also fig.~\ref{fig:S5}(d).
\end{enumerate}
The average spectra after these two corrections are applied were denoted by $\langle \Phi_{(A,B,+,-)}^2\rangle_{\rm corr}$. We note that, to avoid Savitzky-Golay filter oscillations at bin edges, the Savitzky-Golay filtering was performed using data points within each bin plus data points in buffer regions of width $200\beta^2\nu_b$ on both lower and higher frequency edges~\cite{Brubaker2017a}.

In order to search for an axion signal, the data within each frequency bin were optimally filtered by convolving with the expected axion lineshape (\ref{eq:100}). The separation between distinct axion search frequencies was set to $\beta^2\nu_b/2$~\cite{Brubaker2017a,Foster2018}. The optimally filtered PSD data were denoted by $\langle \Phi_{(A,B,+,-)}^2\rangle_{\rm corr,of}$. A histogram of these data points was formed and tested for normality using the Lilliefors test. If the test confirmed that the distribution was normal, its standard deviation $\sigma$ was found by fitting to the Gaussian model. The candidate detection threshold was set to $3.355\sigma$, equivalent to 95\% confidence interval for a $5\sigma$ detection~\cite{Brubaker2017a}. All points above the $3.355\sigma$ threshold were flagged as candidates. For some bins, the Lilliefors test indicated a deviation from normality. For these bins, the distribution was modeled as a sum of two Gaussians, and the candidate detection threshold was set to $3.355\times$ the standard deviation of the broader of the two distributions.

In the 3~kHz to 3~MHz frequency range, with our analysis parameters, there are 20043605 possible ALP frequency points. Assuming the PSD spectra are normally distributed, we would statistically expect 7954 candidate frequencies, where the PSD value is above the $3.355\sigma$ threshold in any given channel. Our analysis procedure produced 16292 candidates in channel $\langle \Phi_A^2\rangle_{\rm corr,of}$, 35375 candidates in channel $\langle \Phi_B^2\rangle_{\rm corr,of}$, 27278 candidates in the symmetric combination channel $\langle \Phi_+^2\rangle_{\rm corr,of}$, and 12513 candidates in the antisymmetric combination channel $\langle \Phi_-^2\rangle_{\rm corr,of}$. The extra candidates were due to residual RF interference and lower-frequency vibrational tones that got past our spectral correction procedure.

Resonant experiments, such as ADMX and HAYSTAC, perform re-scans of candidate frequencies to check if they are statistical deviations or a real axion signal. In our analysis, we used the two axion search channels to make this decision. At each candidate frequency, we evaluated the following logical values:\\
(C1) Is this candidate above the threshold in channel $\langle \Phi_-^2\rangle_{\rm corr,of}$?\\
(C2) Is this candidate below the threshold in channel $\langle \Phi_+^2\rangle_{\rm corr,of}$?\\
(C3) Is this candidate above the threshold in channel $\langle \Phi_A^2\rangle_{\rm corr,of}$?\\
(C4) Is this candidate above the threshold in channel $\langle \Phi_B^2\rangle_{\rm corr,of}$?\\
An axion-like dark matter detection claim would be made for the candidates that satisfy all four conditions. This would correspond to the signals that appear in both SQUID channels A and B and in the antisymmetric combination, but not in the symmetric combination. None of the candidates satisfied all four of our ALP detection conditions. 
Therefore, for each frequency bin, we quoted the ALP coupling value that corresponds to the $5\sigma$ value of the magnetic flux PSD for that bin calculated using eq.~(\ref{eq:Phi_a}). This corresponds to the 95\% confidence interval limit on the axion coupling strength for that bin.

Table~\ref{tab:S6} shows the uncertainty budget for the parameters that were used to calculate the axion-like dark matter coupling limits using eq.~(\ref{eq:Phi_a}). The total uncertainty was dominated by the calibration of the SQUID detector transfer function discussed in section~\ref{sec:squidcal}. We emphasize here again that, when calculating the $g_{a\gamma}$ limits, we used the conservative calibration parameters obtained with the central calibration loop. Therefore, the final uncertainty is not symmetric but extends in the direction of stronger $g_{a\gamma}$ limits.

\begin{table}[t!]
    \centering
    \caption{Sources of uncertainty in~$|g_{a\gamma}| a_0$.}
    \label{tab:S6}
    \begin{tabular}{lc}
        \hline\hline
        Source & Uncertainty \\
        \hline
        Internal magnetic field~$B_0$ & 2\% \\
        Effective volume~$V$ & 4\% \\
        SQUID detector transfer function~$x$ & 25\% \\
        Total uncertainty & 25\% \\
        \hline\hline
    \end{tabular}
\end{table}

\section{Testing the data analysis procedure with simulated ALP signals}

\begin{figure}[h!]
	\centering
	\includegraphics[width=\textwidth]{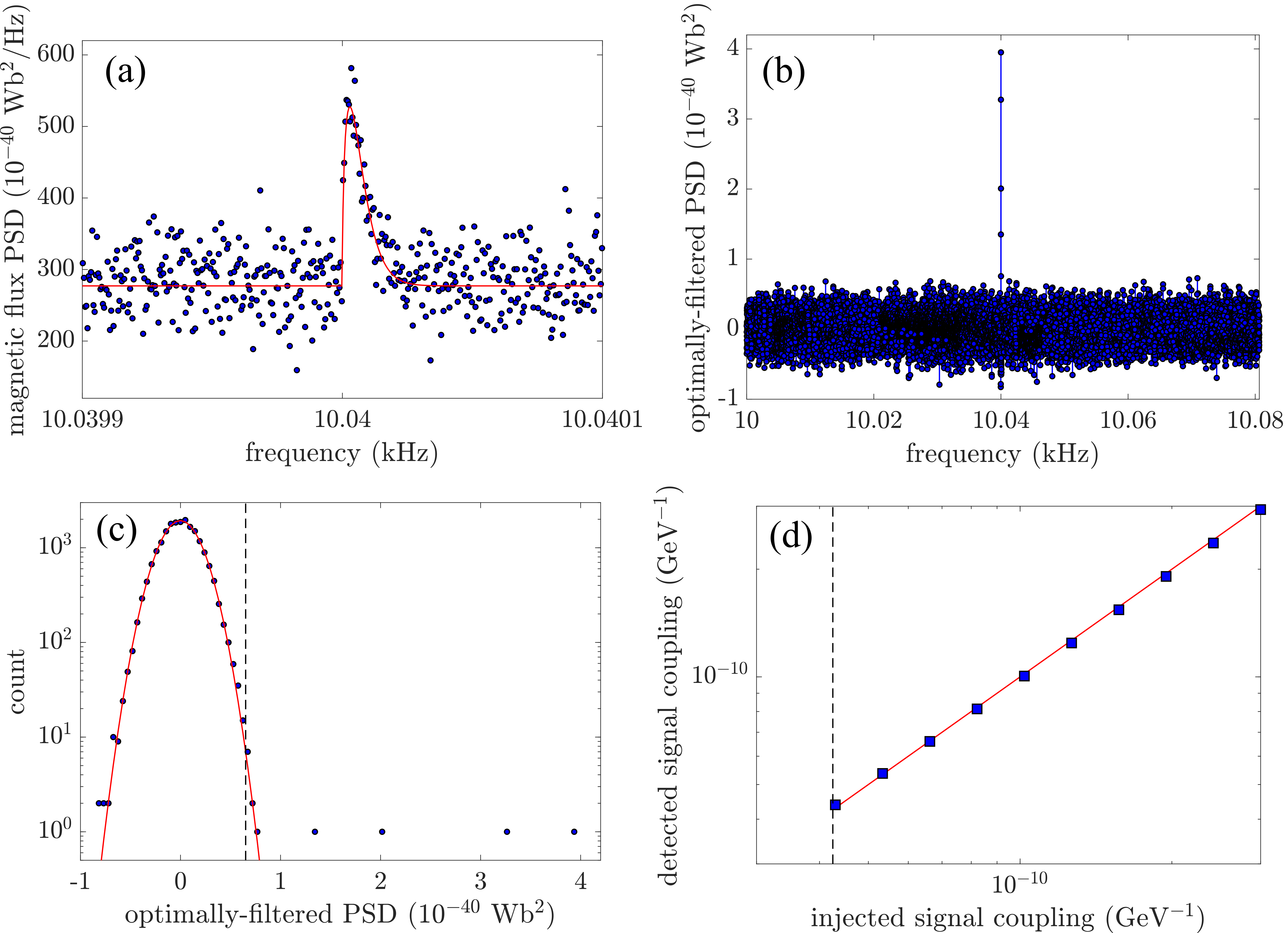}
	\caption{Injecting simulated axion-like signals into the experimental data. (a) A small window of the experimental power spectrum, with a numerically-generated synthetic axion-like signal added (blue points). The synthetic ALP Compton frequency was set to 10.04~kHz and the coupling strength was set to $10^{-10}$~GeV$^{-1}$. The red line shows the standard halo model lineshape, eq.~(\ref{eq:100}). (b) The optimally-filtered PSD, after spectral corrections and convolution with the standard halo model lineshape. Data are shown for frequencies in a single bin, centered near 10.04~kHz. (c) The histogram of the optimally-filtered PSD data within this bin (blue points), and a Gaussian fit (red line). The vertical dashed line marks the $3.355\sigma$ threshold for an event candidate. The injected synthetic axion-like signal was detected with $20\sigma$ significance. (d) Verification of detection and coupling strength recovery of 100 synthetic axion-like signals, injected at random frequencies, with coupling strengths between $10^{-11}$~GeV$^{-1}$ and $3\times 10^{-10}$~GeV$^{-1}$. The coupling strengths recovered from detected signals are shown as blue squares, and the coupling matching the injected signal is shown as the red line. The vertical dashed line marks the $3.355\sigma$ detection threshold, equal to $4.3\times 10^{-11}$~GeV$^{-1}$ for this frequency bin.}
	\label{fig:S5}
\end{figure}

Our data analysis procedure was tested by injecting into our experimental data simulated axion-like dark matter signals that followed the standard halo model with a range of coupling strengths and frequencies between 3~kHz and 3~MHz. The spectral lineshape of such injected ALP at the Compton frequency of 10.04~kHz and with the coupling strength $10^{-10}$~GeV$^{-1}$ is shown in fig.~\ref{fig:S5}(a). The corresponding corrected and optimally-filtered magnetic flux PSD data are shown in fig.~\ref{fig:S5}(b), where the ALP signal at 10.04~kHz is clearly visible. The histogram of these data is shown in fig.~\ref{fig:S5}(c), together with the Gaussian fit and a vertical dashed line marking the $3.355\sigma$ threshold for an event candidate. For this coupling strength, the simulated ALP signal is detected with $20\sigma$ significance.

The detection and recovery of the ALP coupling strength was checked by injecting 100 synthetic ALP signals at coupling strengths between $10^{-11}$~GeV$^{-1}$ and $3\times 10^{-10}$~GeV$^{-1}$. The coupling strengths recovered from detected signals are shown as blue squares in fig.~\ref{fig:S5}(d), and the coupling matching the injected signal is shown as the red line. The vertical dashed line marks the $3.355\sigma$ detection threshold, equal to $4.3\times 10^{-11}$~GeV$^{-1}$ for this frequency bin (centered at 10.04~kHz). 

\bibliography{library}